\documentclass[mathleft
]{an}
\usepackage{graphicx}
\usepackage{times}
\begin{document}

\Pagespan{789}{}
\Yearpublication{2006}%
\Yearsubmission{2005}%
\Month{11}%
\Volume{999}%
\Issue{88}%

\title{Thread Safe Astronomy}

\author{Rob Seaman\inst{}\fnmsep\thanks{Corresponding author:
  \email{seaman@noao.edu}\newline}}
\titlerunning{Thread Safe Astronomy}
\authorrunning{Rob Seaman}
\institute{National Optical Astronomy Observatory,
950 N. Cherry Ave., Tucson, AZ, USA}

\received{}
\accepted{}
\publonline{later}


\abstract{Observational astronomy is the beneficiary of an ancient
chain of apprenticeship.  Kepler's laws required Tycho's data.  As the
pace of discoveries has increased over the centuries, so has the cadence
of tutelage (literally, "watching over").  Naked eye astronomy is
thousands of years old, the telescope hundreds, digital imaging a few
decades, but today's undergraduates will use instrumentation yet unbuilt --
and thus, unfamiliar to their professors -- to complete their doctoral
dissertations.  Not only has the quickening cadence of astronomical
data-taking overrun the apprehension of the science within, but the
contingent pace of experimental design threatens our capacity to learn
new techniques and apply them productively.  Virtual technologies are
necessary to accelerate our human processes
of perception and comprehension to keep up with astronomical instrumentation
and pipelined dataflows.  Necessary, but not sufficient.  Computers can
confuse us as efficiently as they illuminate.  Rather, as with neural
pathways evolved to meet competitive ecological challenges, astronomical
software and data must become organized into ever more coherent `threads'
of execution.  These are the same threaded constructs as
understood by computer science.  No datum is an island.}

\maketitle

\section{Preface}

The title case study in Oliver Sacks' {\em The Man Who Mistook His Wife
for a Hat} (Sacks 1998) concerns a gifted musician facing -- not
blindness -- but a deficit in perceptive ability:

\begin{quotation}
{\em ...not only did Dr P. increasingly fail to see faces, but he saw
faces when there were no faces to see}
\end{quotation}

Dr P's cognition and sight itself were unaffected and he had no
trouble recognizing geometric forms:

\begin{quotation}
{\em `A dodecahedron, of course.  And don't bother with the others -- I'll
get the icosahedron, too.'}
\end{quotation}

Complex objects defied description, however.  The patient's
ailing right hemisphere blocked comprehension, creating
a blindness of {\em knowing}, not of seeing:

\begin{quotation}
{\em He took [a red rose] like a botanist or morphologist given a
specimen, not like a person given a flower.

`About six inches in length,' he commented.  `A convoluted red form with
a linear green attachment.'

`Yes,' I said encouragingly, `and what do you think it {\em is}, Dr P.?'

`Not easy to say.'  He seem perplexed.  `It lacks the simple symmetry of
the Platonic solids, although it may have a higher symmetry of its
own... I think this could be an inflorescence or flower.'

`Could be?' I queried.  `Could be,' he confirmed.

`Smell it,' I suggested, [He] took it to his nose.  Now, suddenly,
he came to life.

`An early rose.  What a heavenly smell!'

Reality, it seemed, might be conveyed by smell, not by sight.}
\end{quotation}

Handed a glove, Dr P. offered no glimmer of recognition, but rather
could only pursue a purely intellectual exploration of the anonymous
object:

\begin{quote}
{\em `A continuous surface, [...] infolded on itself.  It appears to have' --
he hesitated -- `five outpouchings, if this is the word.'}
\end{quote}

It was only later upon donning the glove that Dr P. exclaimed,
`My God! It's a glove!'  As Sacks puts it:

\begin{quotation}
{\em [T]he right hemisphere [of the brain] controls the crucial powers of
recognising reality [...]  The left hemisphere, like a computer [is]
designed for programs and schematics [...]

[Dr P.] construed the world as a computer construes it, by means of key
features and schematic relationships.  The scheme might be identified --
in an `identi-kit' way -- without the reality being grasped at all.}
\end{quotation}

The original Identi-Kit{\small \textregistered} was marketed to the law
enforcement community by Smith \& Wesson (1968) as a tool for constructing
facial composites from innumerable disjoint details related by observers
(witnesses).  Facial compositing is now often done with software, and
in fact, an astronomical software package for matching schematic simulations
(scientific models) to observations has previously been compared to
Identi-Kit{\small \textregistered} (Hibbard 2004).  As with the police
sketch artist's rendering of a suspect -- a rendering compiled second-hand
from witnesses' conflicting recollections -- so do astronomers apprehend
the universe.

\section{Through a Glass Darkly}

Of all the sciences, astronomy is most dependent on what can be seen
at a distance.  Contraints such as field of view, resolving power, bandpass,
signal/noise ratio, instrumental artifacts, relative motion, light travel
time, foreground obscuration, and background emission -- to name just a
few -- can confound our understanding of even familiar phenomena.
Astronomers, however, are rarely faced with a task equivalent to
identifying a commonplace flower or glove.  The objects of our
regard are the most mysterious in the universe, demanding much
from our earthbound imaginations.

Overtly similar phenomena -- e.g., types of supernovae -- may be
attributed to vastly different physical events.  Alternately, identical
processes occurring to similar objects may appear very different
depending on the context.  Relatively simple objects may only be revealed
through mind boggling means such as gravitational lensing.  Other objects
and processes are forever hidden by the cosmic censor.

If astronomical perception is to rise above Sacks' stultified
`identi-kit', a coherent strategy must be followed of tying one
observation (the astronomical counterpart of a sight, smell, touch
or other human sensation) to other observations in a lucid
gestalt - a coordinated workflow.  There is a name for this in
computer science:  an execution `thread'.  See, for example,
Oaks \& Wong (2004).

\section{Threaded architectures}

One can therefore state the {\em First Law of Observing}, namely: all
observations are part of an empirical thread.  This is true whether
or not loose ends unravel, and is true both for time
domain and static phenomena.  Each observation requires context from
earlier data-driven studies, and each observation provides context
for steering future investigations.

Moreover, the meaning of `context' here is precisely how the word is
used in computer science.  The full state (values of all variables,
pointers, data structures, and program instructions) required to
reconstitute a running program (software or observational) is
efficiently saved such that other contexts can be swapped back
in for running.

\subsection{Multi-threading}

One obvious aspect of a threaded architecture is that it supports
multiple threads.  All telescopes are scheduled in some fashion.  It may
be possible to say that all astronomical facilities -- whether
empirical, theoretical, virtual, or otherwise -- must be scheduled.  What
is being scheduled in each case are alternative threads of execution,
shadowed by a thread of some sort of data products or science products.
Examples include competatively allocated, block scheduled, telescope
time as provided by various national observatories.  Threading is perhaps
even more evident with queue scheduled facilities, where each project in
an interleaved queue is serviced in turn as contingent conditions are
encountered for each prioritized observing program.

Features of threaded computer science architecture map directly to a
threaded astronomical (or in general, empirical) architecture.
Lightweight and simultaneous observing tasks are pursued
through time slicing of forked or replicated
astronomical contexts.

\subsection{Process safety}

Lightweight threads are distinct from heavyweight process
architectures, but a number of concepts apply at least by analogy.  The
execution of computer processes is governed by a scheduler common
to all.  Every aspect of the context of each process is controlled (as
should be true of telescope and camera contexts).  For example, this permits
context switching at almost any point during execution, except
during atomic system calls.  Contexts can be swapped out to secondary
storage, perhaps as a form of virtual memory -- that is, common system
facilities such as memory can themselves benefit from threaded handling.
Memory is protected such that misbehavior in one context cannot sabotage
another.  Processes can interact in various ways through interprocess
communication (IPC) protocols.

\subsection{Thread safety}

By contrast, threads operate with shared resources, such as a 
common address space.  As with computer science architectures, what is
required are {\em thread safe} astronomical systems.  Provisions are
required in our software interfaces, hardware implementations, and
staff procedures to manage concurrency in access to resources of
all types, i.e., semaphores are used in applications to avoid race
conditions and permit reentrancy.  Translation: investigators must
communicate with each other to avoid demanding the same resources at
the same time, often by interrupting one observing program with another.

Better yet, however, would be to embed such protections in astronomical
systems so as to apply to all observatory activities and thus all threads
of empirical context and data.

\section{Threads in autonomous astronomy}

The challenge conceiving autonomous systems for astronomical purposes
(Seaman et al. 2007) is not to build a single robotic telescope,
rather it is to build a complete networked ecosystem of automated
robotic assets like those of the Heterogeneous Telescope Networks
(HTN) consortium (Allan et al. 2006).  This can be extended by recognizing
that automation, per se, is a convenience, not a necessity.  Human
mediated, as well as robotic, assets can form a procedural `system
of systems' (Humphrey 2006) for accomplishing astronomical goals
(Smith, Seaman \& Warner, 2008).

\subsection{Some issues for an astronomy ecosystem}

A prerequisite for any coordinated scientific activity is a reliance
on robust technical and logistical standards.  For astronomy,
in addition to various familiar standards of the International
Astronomical Union pertaining to nomenclature, for instance, or to
software, e.g., FITS (Wells, Greisen, \& Harten 1981), it is now clear
that future developments will feature standards of the International
Virtual Observatory Alliance (IVOA) -- see for example, Brunner,
Djorgovski, \& Szalay (2001) -- coordinated under Commission 5 of the IAU.

Observatory scheduling policies, new observing modes (Boroson, Davies,
\& Robson 1996) and other operational paradigms have a profound influence
on how the resulting data products may be used.  Data from a classically
scheduled telescope will be more heterogeneous than data from a queue
scheduled telescope, and both more so than from a robotic facility.
Calibration data may be non-standard or missing.  Metadata may require
the observing logs to interpret.  Those observing logs may be unavailable,
even if the data are archived.  However, a fully robotic telescope may
not permit unplanned observing sequences or the exploration of creative
new instrumental methods.

In a sense, remote observing, or telepresence, (Emerson \& Clowes
1993) embodies an opposing suite of technologies from those employed by
robotic telescopes.  The observer's influence is not automated, but
extended from the mountaintop.  These are not incompatible,
however, but rather the question is how to optimize autonomous
activities when these are called for, and how to optimize human
influence when this is appropriate.

A completely proprietary observation means cutting the empirical
thread -- severing one observation from those that might otherwise
follow from other investigators.  No single observation can stand alone.
Each benefits from the rich investigational fabric, community woven.
An absolute policy of public data rights, however, is unsustainable in
a research environment dependent on competitive funding and oversubscribed
observing time.  Many observatories rely on a finite proprietary period
for most data products, with a mixture of immediate availability for some
(e.g., transient alerts) and embargoed access for others (e.g.,
dissertation data).

Astronomy has long functioned in a medieval fashion of bartering
observing opportunities and thus data on the one hand -- and of
treating these as state secrets on the other.  Some are contemplating
developing an astronomical market economy (Etherton, Steele, \&
Mottram 2004).  Care must be taken to ensure that observations resulting
from such pay-per-view modes of operation are properly curated and
interpreted in their complete scientific context.  Control Theory
(Doyle, Francis, \& Tannenbaum 1990) may supply a well-needed
counter balance to free market chaos.

Data must be portable (in all senses of the word) to express their full
value. Data transport (Seaman et al., 2005) is thus of key concern in
preserving and
extending connections from earlier observations to later ones.  A
single archival observation may intersect many threads:  be used
for multiple purposes and benefit multiple investigators.  As each data
product is transported, so must its links and metadata.

\subsection{Implementation independence}

Note is often taken of the technological choices of some new facility
in implementing component subsystems.  Rather,
the choice of solving a particular problem in hardware or software -- or
even `bioware' (Bahill \& Dean 2007) -- is a detail like all others to be
traded off against requirements.  The art of system design is in
recognizing that the system has an existence beyond its prototypes.
Tools such as the Unified Modeling Language (Fowler 2004) can aid in
comprehending the features of a system independent of technical
distractions.

For example, the trade-off between hardware, software, and bioware
(procedural recipes requiring staff or user actions) often
depends on real-world issues like the necessity of interoperating with
legacy systems, or the availability of commercial-off-the-shelf (COTS)
solutions.  Hardware options may be
state of the art, but expensive as a consequence.  Software, on the
other hand, is often regarded as inexpensive and flexible.
The minimal up-front software costs may well be balanced, however, by
increased maintenance.

%
%

\subsection{An example of threading: the VOEvent lifecycle}

The IVOA VOEvent standard (Seaman et al. 2006) implements a
publish/subscribe mechanism for conveying alerts of transient
celestial events (Williams \& Seaman 2006).  It is the nature of
such alerts to generate follow-up observations and thus a sequence
of additional alerts.  VOEvent provides a citation mechanism for
connecting rich threads of such follow-ups, including logistical
options such as tying two threads into one, dividing one thread in
two, or retracting an alert and thus cutting the thread.  References
embedded in VOEvent packets link to archival data and to dynamic
web services, extending threads further.

It is these threads of data products that convey (Seaman \& Warner
2006) and create telescope behavior through execution threads of the
telescope and instrument interfaces of robotic environments such as
HTN, as well as the traditional computerized and human mediated
environments typical of both classically and service operated telescopes.

\begin{figure*}
\begin{center}
{\includegraphics[width=\linewidth]{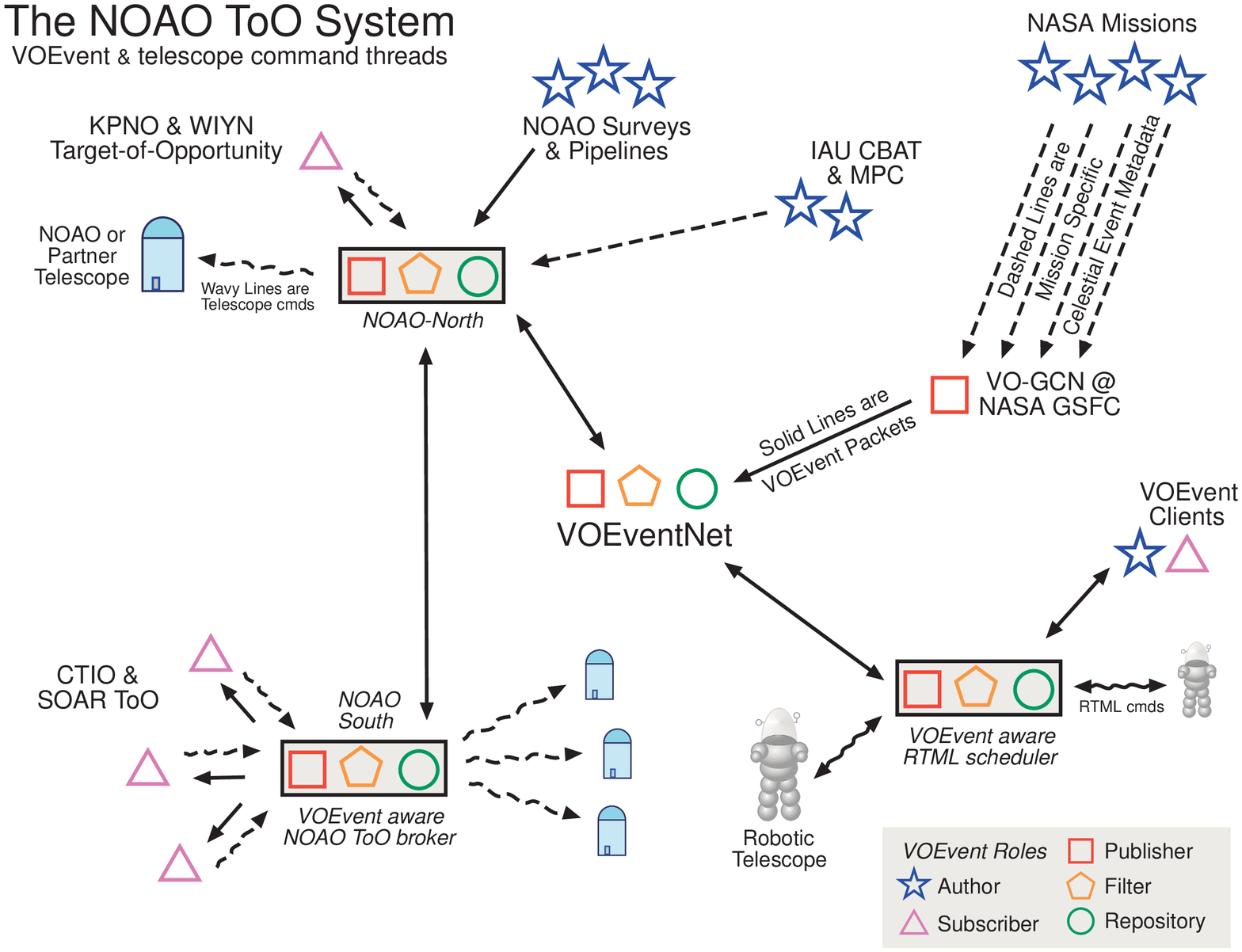}
\caption{The NOAO Target of Opportunity observing system is an example
of threads in action.  Transient alerts in the form of VOEvent packets
trigger behavior from observatory facilities and personnel.  In turn,
execution threads in the form of autonomous and human-mediated commands
to telescope and instrument subsystems trigger follow-up observations
and the resulting data products.  Threads from multiple projects and
investigators are adaptively interleaved.  The system, and thus the
need for protected threaded execution, encompasses the observing and
virtual assets of the entire astronomical community.}
\label{ToO_threads}}
\end{center}
\end{figure*}

\section{Summary}

A coherent view of the observing process and of the ubiquitous
threads of astronomical data is required to avoid the
empty understanding of {\em identi-kit astronomy}.  Astronomical
techniques benefit from the realization that these threads are
rigorously analogous to the threads of computer science.


\end{document}